\documentclass[showpacs,amssymb,floatfix,pre,aps,notitlepage]{revtex4-1}

\usepackage{amssymb,amsfonts,bm}
\usepackage{graphics,graphicx}
\usepackage{psfrag}

\usepackage{color}

\newcommand{\emma}[1]{#1}

\begin{document}

\title{Mean-field beyond mean-field: the single particle view for moderately to 
strongly coupled charged fluids}

\author{Ladislav \v{S}amaj}
\affiliation{Institute of Physics, Slovak Academy of Sciences, Bratislava, Slovakia}
\author{Alexandre P. dos Santos}
\author{Yan Levin}
\affiliation{Instituto de F\'{i}sica, Universidade Federal do Rio Grande 
do Sul, CP 15051, CEP 91501-970 Porto Alegre, RS, Brazil}
\author{Emmanuel Trizac}
\affiliation{LPTMS, CNRS, Univ. Paris-Sud, Universit\'e Paris-Saclay, 91405 Orsay, France}

\begin{abstract}
In a counter-ion only charged fluid, Coulomb coupling is quantified by a single dimensionless parameter. 
Yet, the theoretical treatment of moderately to strongly coupled charged fluids is a difficult task,
central to the understanding of a wealth of soft matter problems, including biological systems.
We show that the corresponding coupling regime can be remarkably well described by a
single particle treatment, which, at variance
with previous works, takes due account of inter-ionic interactions. 
To this end, the prototypical problem of a planar charged dielectric interface is worked out. Testing 
our predictions against Monte Carlo simulation data reveals an excellent agreement.
\end{abstract}


\date{\today}

\maketitle

\section{Introduction}
Charged fluids are abundant in man-made or natural systems, in which thermalized mobile ions interact via
Coulomb forces collectively, and also with more macroscopic charged bodies such as colloids, proteins, or DNA.
The first theoretical attempt for describing inhomogeneous Coulomb fluids dates back about a century ago, to pioneering  works 
of Gouy in Lyon \cite{Gouy10} and Chapman in Oxford \cite{Chap13}. 
These predate the Debye and H\"uckel approach which aimed at accounting for
the unusual thermodynamic properties of electrolytes like NaCl, where dissociation leads to a fluid of 
Na$^+$ and Cl$^-$ ions in water \cite{DeHu23}. These early treatments are all mean-field in spirit.
It was realized in the 1980s that by discarding electrostatic correlations, mean-field theory precludes some counter-intuitive 
effects such as the electrostatic attraction of like charge surfaces, revealed
by experiments, simulations, and theoretical approaches, 
see \cite{LiLo99,WBBP99,GBPP00,SodlC01,GrNS02,Levi02,BAO04,NJMN05,NJMN05,Mess09} and references
therein. 
It is now recognized that the validity of mean-field treatments, epitomized by the Poisson-Boltzmann theory
of extensive use in colloid science \cite{Ande06}, 
requires the necessary condition of sufficiently small electrostatic coupling; in
the language of the coupling parameter $\Xi$ to be defined below and which pits electrostatic against thermal energies,
this means $\Xi \ll 1$ up to $\Xi \simeq 1$. 
On the other hand, systems with moderate to strong coupling are profuse, 
starting with nucleic acids and cell membranes in aqueous solutions. 
Charges are pivotal to their stability {\it in vivo}.
The study of these biological objects from a physics perspective has rekindled interest in Coulomb fluids,
with particular emphasis on strong coupling regime.
\emma{Yet, analytical progress for moderately to strongly coupled charged fluids has proven elusive,
as will be illustrated below. Our goal here is to fill this gap, with a theoretical treatment 
that is both physically transparent, and remarkably accurate. It takes advantage of the existence
of a correlation hole around individual ions in the system, a well known feature, that has nevertheless not
been turned into an explicit analytical treatment so far. It is also relevant to emphasize from  
the outset that our approach deals with salt-free systems, 
where only counterions are present in the solution. This situation, with no added buffer electrolyte,
applies to deionized suspensions (see e.g. the experiments reported in \cite{PMGE04}).}

\begin{figure}[htb]
\begin{center}
\includegraphics[width=0.37\textwidth,clip]{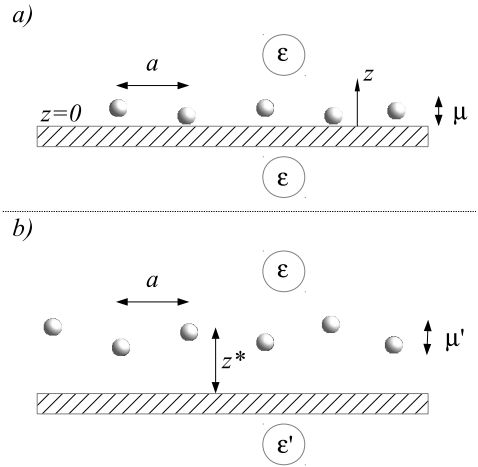}
\caption{Schematic side view of the system, without (panel a) and with (panel b) dielectric mismatch.
The mobile counter-ions, point-like, are drawn 
as spheres for the sake of illustration.
In a), the dielectric constant of the solvent ($\varepsilon$) and that of the
interface ($\varepsilon'$) are equal. We will also consider in b) the case where both constants differ,
for which the dielectric mismatch is quantified by $\Delta=(\varepsilon-\varepsilon')/(\varepsilon+\varepsilon')$.
Panels a) and b) depict regimes of large Coulomb coupling ($\Xi\gg1 $). Then, the characteristic distance $a$
between the counter-ions is set by electro-neutrality: $\sigma a^2 \propto q$, where $\sigma e$ is the plate surface charge
density at $z=0$ and $-qe$ is the ion's charge, with $e$ the elementary charge. 
The typical extension $\mu$ follows by balancing thermal energy $kT$ with the energy of an ion 
$-q e$ at position $z$ in the potential $-2\pi \sigma e z/\varepsilon$ created by the bare plate: 
$\mu = \varepsilon \,kT / (2\pi q \sigma e^2)$, the so-called Gouy length. 
The coupling parameter is defined as $\Xi = 2 \pi \sigma q^3 e^4 /(\epsilon k T)^2$.
Thus, $\Xi \propto a^2/\mu^2$ and $\Xi \gg 1 \Rightarrow \mu \ll a$. 
In panel b), repulsive dielectric images should be considered 
($\varepsilon' < \varepsilon$)
and a depletion zone of size $z^*$ appears.
The typical extension of the profile, $\mu'$, is no longer given by $\mu$ \cite{rque20}.
}
\label{fig:model} 
\end{center}
\end{figure}

\section{Length Scale Separation}

The limit of asymptotically large couplings admits a simple description,
in elementary settings such as that sketched in Fig. \ref{fig:model}-a.
It can be understood by a length scale analysis, which we now illustrate on the emblematic primitive 
counter-ion only model. 
\emma{For strongly charged plates, most  counterions  remain in a close vicinity of 
the surface. The characteristic distance $a$
between the condensed counter-ions is ruled by electro-neutrality: $\sigma a^2 \propto q$, 
where $\sigma e$ is the plate surface charge
density at $z=0$ and $-qe$ is the ion's charge, with $e$ the elementary charge. 
The typical extension, or excursion of the counter-ions from the surface, is denoted $\mu$.
This quantity, named the Gouy length, follows by the balance of 
thermal energy $kT$ with the energy of an ion 
$-q e$ at position $z$, 
$\mu = \varepsilon \,kT / (2\pi q \sigma e^2)$. 
The dimensionless coupling parameter, defined as $\Xi = 2 \pi \sigma q^3 e^4 /(\epsilon k T)^2$, is proportional 
to $a^2/\mu^2$.  
When $\Xi \gg 1$, Coulomb interaction between the 
counter-ion exceeds
thermal energy, so that the mobile counter-ions in the vicinity of a plate
are strongly attracted to the surface, and at the same time repelled from the 
adjacent counterions, $\Xi \gg 1 \Rightarrow \mu \ll a$.  This results in a 
correlation hole size $a$
\cite{RoBl96,Shkl99}, exceeding a typical transverse excursion of a counter-ios from the surface characterized by
the Gouy length $\mu$, see Figure \ref{fig:model}-a, where the key length scales are depicted.  For colloidal particles with 
bare charge $Z=10^4e$ and radius of $R=10^3$ \AA, in aqueous solution, 
the coupling parameter is $\Xi \approx 0.26$ for monovalent counterions ($q=1$),
$2.1$ for divalent counter-ions, and $7.0$ for trivalent counter-ions.  However, since $\Xi$ is inversely proportional to 
the square of the dielectric constant, for solvents of lower dielectric constants such as mixtures containing  
water and alcohol, $\Xi$ can easily reach $50$ for 
moderately charged surfaces with trivalent counterions. 
It is also relevant to provide reasonable bounds for the possible values of $\Xi$, as a function 
of valence $q$. In water at room temperature, highly charged interfaces
have $\sigma e$ on the order of one $e$ per namometer square, 
and therefore $\Xi$ is on the order of $q^3$. With trivalent ions,
this means $\Xi\simeq 30$, which is already way into the regime covered by our treatment.}

The length scale separation provides the grounds for a surprisingly simple picture of a strongly correlated Coulomb system 
where the ions react mostly to the bare plate potential, while ion-ion interactions become insignificant as 
$\Xi\to\infty$ \cite{Shkl99,NJMN05,Mess09}.
\emma{Thus, the ionic density profile
takes an exponential form $\rho(z) \propto \exp(-z/\mu)$ characteristic 
of a particle in a constant field.  The proportionality factor can be 
determined by the contact value 
theorem \cite{HeBl78}.  }
This ``ideal gas'' barometric law has been fully validated by numerical simulations \cite{MoNe02,Varenna}.
Corrections beyond the ideal gas regime can be computed in a $1/\sqrt{\Xi}$ expansion 
by a perturbation around the Wigner crystal \cite{SaTr11}, that forms when $\Xi$ exceeds some (very large) 
crystallization value $\Xi_c \simeq 3 .10^4$ \cite{BaHa80}. 

It is generally believed that single particle ideas fail in situations where scale separation 
no longer holds: for instance if $\Xi$ is in some crossover regime of
moderate coupling or in the situation of Fig. \ref{fig:model}-b) 
with a dielectric mismatch. 
We shall see that although the ideal gas view indeed severely breaks down in these generic cases 
-- which as a matter of fact
significantly limits its practical interest -- a ``correlation hole modified'' single-particle treatment 
can be effectively applied. It is our purpose to present this fully analytical, self-consistent approach.
The theory developed here allows to accurately determine the counter-ion density distribution 
$\rho$, which is in striking agreement with computer simulation results.  This leads to an
unexpected conclusion that somewhat beyond the usual mean-field regime of weakly coupled fluids, 
an even simpler mean-field provides a quantitative description. 
In the limiting cases where the ideal gas formulation is relevant, our analysis recovers it.


\section{Correlation Hole: Treatment and Consequences}

We now address the simplest geometry where lack of scale separation forestalls the ideal gas single particle physics:
the planar interface alluded to above, with a dielectric jump between
the solvent (dielectric constant $\varepsilon$) and the confining charged body (dielectric constant $\epsilon'$)
occupying the lower half space as shown in Figure \ref{fig:model}-b.
Although simplified, such a geometry provides a paradigmatic testbed to shape intuition and theoretical ideas.
The situation $\Delta = (\varepsilon-\varepsilon')/(\varepsilon+\varepsilon') >0 $ is the most relevant one, 
since the dielectric constant of materials like glass, proteins, or polarizable colloids is much smaller than that of water:
each charge admits an image of the same sign \cite{Jackson}, with a resulting repulsive interaction. 
\emma{It also encompasses the air-liquid interfaces, for which $\varepsilon/\varepsilon' \simeq 80$.
The case $\Delta <0$ leads to attractive images \cite{SaTr12CPP}, 
and to the disappearance of the depletion zone in 
Fig. \ref{fig:model}-b. The extreme limit corresponds to a grounded electrode with $\varepsilon' \rightarrow \infty$
for which $\Delta = -1$.  In this case the ions can no longer be modeled as point particles and a hardcore must be introduced.
In this paper we will restrict our attention to systems with  $\Delta>0$.}

The mobile ions are attracted to the oppositely charged interface at $z=0$, but concomitantly 
each charge $-qe$ at position $z$ has a dielectric image of charge $-qe \Delta$ at $-z$ \cite{Jackson},  
which strongly repels it. A depletion zone ensues \cite{OnSa33};
it is quite straightforward to estimate its size $z^*$, which turns out to be of the same order as $a$.
Thus, one can no longer consider that ions are far from each other compared to their distance to the plate: 
the intrusion of a new length scale, $z^*$, explains the failure of the single particle ideal gas picture. 
Nevertheless, the ionic profile's extension, $\mu'$, remains the smallest length scale of the problem \cite{rque20}. 
Hence, we are led to neglect the correlations between the ion's fluctuations, while taking due account 
of their interactions in an effective way, at variance with the ideal gas formulation.
The problem we face reduces to computing the effective potential $u$ that a given ion experiences, when at a distance 
$z$ away from the interface. When known, $u$ directly leads, through a Boltzmann weight, to the main quantity 
of interest, the density profile: $\rho(z) \propto \exp(-\beta u) $, $\beta=1/(kT)$ being the inverse temperature.
We emphasize that when explicit analitic expressions are sought, the state 
of the art lies in the single particle ideal gas view, in which case the potential of mean-force $u$ stems from the force due to the plate 
at $z=0$ and the test particle image charge \cite{NJMN05,Mess09,MoNe02,MoNeEPL02}. 
We shall see that this treatment is inappropriate for $\Delta \neq 0$, so that there
is no analytical treatment available in the literature to study this general case. 
We attempt here to fill the gap. 
In other words, while the idea of correlation holes in more or less correlated Coulombic fluids is not novel
\cite{Nord84,RoBl96,Shkl99,BaDH00,GrNS02,DoRu03,BAO04,HaLu10}, transforming the corresponding insight into a fully analytical theory is
new; it is the subject of our paper.

\begin{figure}[htb]
\psfrag{AA}{$z^*$}
\psfrag{BB}{$(1+\Delta) \sigma\,e$}
\psfrag{CC}{$-\sigma\,e$}
\psfrag{DD}{$-\Delta\,\sigma\,e$}
\psfrag{EE}{$2 R_6$}
\psfrag{FF}{$z$}
\psfrag{GG}{$z=0$}
\begin{center}
\includegraphics[width=0.37\textwidth,clip]{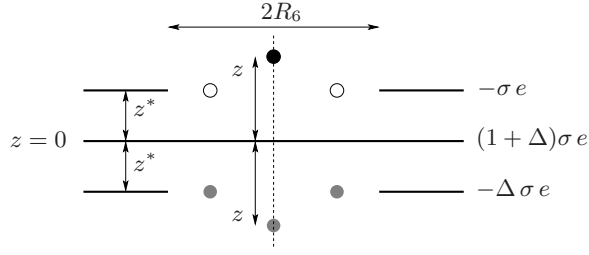}
\caption{Schematics of the ch$_6$ approach. A test particle (filled disc) is singled out at elevation $z$.
Other counter-ions are assumed to be at their typical location $z^*$.
Upon smearing out the counter-ions beyond 
a cutoff distance $R_6$, one obtains a {\em punctured} plate with charge density $-\sigma e$. The empty circles
stand for the 6 nearest neighbors of the test particle. The symmetrically located dielectric images
-- discrete (displayed in gray) or continuous -- are also shown. The simplified 
ch$_0$ view leads to a very similar setup, with the difference that there are no discrete neighbors: 
these ions are also smeared out, so that the hole becomes smaller, 
of radius $R_0 = R_6/\sqrt{7}$.
}
\label{fig:viewch6} 
\end{center}
\end{figure}

Since practically relevant values of the coupling parameter are orders of magnitude smaller than
the crystallization threshold, we envision the ions as forming a liquid, essentially two dimensional
since we do not aim at covering the limit of too small $\Xi$ (we will address the range 
$\Xi>10$ here \cite{rque90}). The key structural features of this liquid are embodied in the
pair correlation function $g(r)$ \cite{HaMD86,rque30}, a function
of inter-ion distance providing the density of neighbors. This $g(r)$ is more or less
structured depending on the value of $\Xi$ \cite{MoNe02}, but is always strongly depleted at small 
distances $r$ due to the strong Coulomb repulsion \cite{Nord84,RoBl96,Chen06,Sant06,HaLu10,BdSL11}: 
we recover the correlation 
hole depicted in Fig. \ref{fig:model}. A second characteristics is that the size of
this hole is essentially $\Xi$-independent: being set by electro-neutrality, 
it is always given by the length scale $a$ introduced in the caption of
Fig. \ref{fig:model} \cite{MoNe02}; besides, each particle has a coordination six 
\cite{rque50}. We claim that these gross features are sufficient for a proper account of the ionic profile,
without inclusion of further details.
Two levels of simplification will be provided, having in common the existence of a
correlation hole around the test particle, in the form of a concentric disk.
1) Apart from the test particle, the fluid of counter-ions is assumed structureless beyond $R_0$ (meaning $g(r)=1$ for $r>R_0$).
The size of the hole is set by balancing the hole and ion charges: $\pi R_0^2 \sigma e = q e$.
This leads to a system of a moving ion
in the field of a plate at $z=0$, a punctured plate at $z^*$ having a circular hole of size $R_0$,
plus the dielectric images of all charges, of the same sign but weighted with a prefactor $\Delta$, and 
located at the symmetric position with respect to the mid plane at $z=0$. 
We call this route the correlation hole + strong coupling with zero neighbor (ch$_0$).
2) In a refined approach, we set $g(r)=1$ beyond the first neighbors. 
Then, each particle with its 6 neighbors is in the center of a hole with radius $R_6$, 
now such that $\pi \sigma R_6^2 = q + 6q = 7q$. 
Due account of image charges leads to the model represented in Figure \ref{fig:viewch6},
referred to as ch$_6$. For both ch$_0$ and ch$_6$ routes, the process of smearing out 
an infinite number of counter-ions leads to a punctured charged plate, with a hole concentric with the
test ion. Its interaction with the test particle is essential for a good account of the density profile.

\section{Results}

\emma{To explore the range of validity of the theory all the results will be compared with the Monte Carlo simulations
performed using the 3D Ewald summation with a correction for slab geometry and for surface polarization. 
More details regarding simulations can be found in Refs.~\cite{DoLe14} and ~\cite{DoLe15}.  An interested reader
can also consult an efficient
implementation of slab geometry simulations for charged interfaces which has 
recently been developed in Ref. \cite{DoGi16}.}

\begin{figure}[htb]
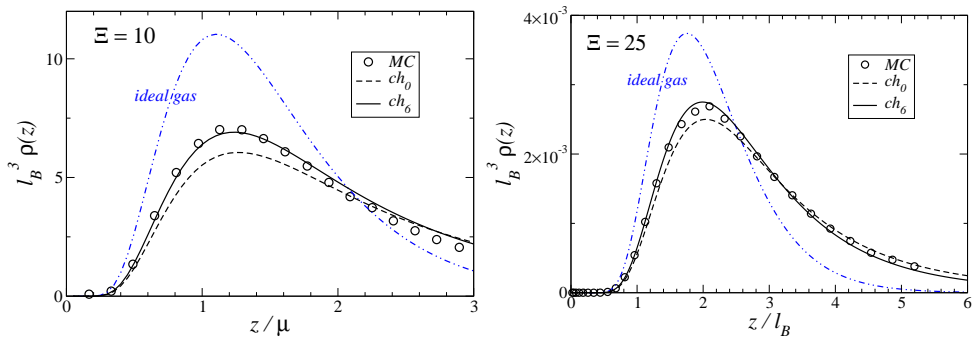

\begin{center}
\includegraphics[width=0.35\textwidth,clip]{rho_Xi10_del0.95_simp.eps}
\includegraphics[width=0.36\textwidth,clip]{rho_Xi25_del1_q5.eps}
\caption{Density profile of counter-ions for $\Delta=0.95$ (meaning $\varepsilon/\varepsilon' \simeq 40$),
$\Xi=10$ (upper graph) and for $\Delta=1$, $\Xi=25$ (lower graph). The ch$_0$ and ch$_6$ predictions
are compared to the ideal gas profile proposed in \cite{MoNeEPL02}, and to the results of
Monte Carlo simulations (taken from \cite{MoNeEPL02} for the upper graph). Here, $l_B=\beta e^2/\epsilon$
is the Bjerrum length.}
\label{fig:smallXi} 
\end{center}
\end{figure}

The analysis now proceeds in two steps \cite{rque97}. First, the optimal
distance $z^*$ is derived, which yields the maximum of the ionic profile $\rho(z)$.
Second, the  effective one-particle potential $u$ is computed. 
\emma{For the sake of simplicity, we start by presenting the ch$_0$ approach. We fix all ions at $z=z^*$ 
(including the test particle), and calculate $E_0$,
the energy per particle of the system, made up of 3 charged planes, two of which are punctured and located 
at $\pm z^*$, and 2 discrete charges (image included). It proves convenient to add and subtract to the image
plane at $z=-z^*$, the potential of a charged disc with same density as the plate, $-\Delta \sigma e$.
In doing so, one obtains a non-punctuated plate at $z=-z^*$, and a disc of charge density $\Delta\sigma e$,
with radius $R_0$. The resulting energy per particle is}
\begin{eqnarray}
E_0(z^*) &=& \frac{2\pi}{\varepsilon}\,(1+\Delta)\,\sigma q \,e^2 z^* \, - \frac{1}{2}\,\Delta\,\frac{2\pi}{\varepsilon}\,\sigma q \,e^2 (2z^*)
\,-\,\frac{1}{2} \Delta q \sigma\frac{e^2}{\varepsilon}\,\int_0^{R_0} dr \,\frac{2\pi r}{\sqrt{r^2+(2z^*)^2}} 
+\frac{q^2 e^2}{2\varepsilon} \Delta \frac{1}{2z^*}
\nonumber\\
&=& 
\frac{2\pi}{\varepsilon}\,\sigma q \,e^2 z^* + \frac{q^2 e^2}{2\varepsilon} \Delta \frac{1}{2z^*}\,
-\, \pi \Delta \,q \,\sigma \frac{e^2}{\varepsilon}\left[\sqrt{R_0^2+(2z^*)^2}-2z^* \right] .
\end{eqnarray}
Turning to the ch$_6$ case, we have to consider 3 charged planes, two of which are punctured and located 
at $\pm z^*$, and 14 discrete charges. \emma{Proceeding along similar lines as above,} the energy per particle now reads:
\begin{eqnarray}
E_0(z^*) =&&\frac{2\pi}{\varepsilon}\,\sigma q \,e^2 z^* +\frac{q^2 e^2}{2\varepsilon} \Delta 
\left[\frac{1}{2z^*} + \frac{6}{\sqrt{a^2 + 4 z^{*2}}}
\right]
\nonumber\\
&&- \pi \Delta \,q \,\sigma \frac{e^2}{\varepsilon}\left[\sqrt{R_6^2+(2z^*)^2}-2z^* \right] .
\end{eqnarray}
Introducing the dimensionless variable $t=2z^*/a$ where $a=3^{-1/4} \sqrt{2q/\sigma}$
\cite{rque60} and minimizing $E_0$ with respect to $t$, we have to solve
$$
 1-\Delta\left[\frac{t}{\sqrt{(R_6/a)^2+t^2}} -1 
\right] = \frac{\sqrt{3}}{4\pi} \left[\frac{\Delta}{t^2} \,+ \,\frac{6\,\Delta \,t}{\left(1+t^2\right)^{3/2}} 
\right].
$$
Once $t$ and thus the depletion zone extension $z^*$ is found, we have to dissociate the test 
particle from the ionic layer, move it along the $z$  axis as depicted in Fig. \ref{fig:viewch6}, 
and compute the resulting
potential $u(z)$. This is another elementary electrostatics exercise \cite{rque70}, with the result:
\begin{eqnarray}
&&\beta u(z) \,= \,(1+\Delta) \, \widetilde z  \,+\, \frac{\Xi \Delta}{4 \, \widetilde z}
\nonumber\\
&& - \,\sqrt{ \left(\widetilde R_6\right)^2 + \left(\widetilde z - \widetilde z^*\right)^2 } \,-\,
\Delta \sqrt{ \left(\widetilde R_6\right)^2 + \left(\widetilde z + \widetilde z^*\right)^2  }
\nonumber\\
&& + \frac{6 \,\Xi}{\sqrt{ \widetilde a^{\,2} + \left(\widetilde z - \widetilde z^*\right)^2 }}
\,+\, \frac{6 \,\Xi \,\Delta}{\sqrt{ \widetilde a^{\,2} + \left(\widetilde z + \widetilde z^*\right)^2 }}
\label{eq:uch6}
\end{eqnarray}
where tilde distances are rescaled by the Gouy length, e.g. $\tilde z = z/\mu$.
The ch$_0$ counterpart of Eq. (\ref{eq:uch6}) is again very similar, without the last two terms 
in $6\,\Xi$ and with the substitution $\widetilde R_6\to \widetilde R_0$ for the hole size. Since $\widetilde R_6^2=14\,\Xi$, we have
$\widetilde R_0^2=2\,\Xi$. Finally, the suitably normalized Boltzmann weight is the density
profile sought for:
\begin{equation}
\rho(z) \,=\, \frac{\sigma}{q}\, \frac{e^{-\beta u(z)}}{\int e^{-\beta u(z')} dz'} .
\label{eq:rho_u}
\end{equation}
By accounting solely for the interaction with the plate at $z=0$ and with the test particle image,
one has $\beta u = (1+\Delta) \, \widetilde z  \,+\, \Xi \Delta/(4 \, \widetilde z)$, which,
when inserted into Eq. (\ref{eq:rho_u}), leads to the ideal gas profile proposed in 
\cite{MoNeEPL02,JKNP08}. Such an approach is expected to fail as soon as the afore discussed scale separation is violated,
that is whenever $\Delta \neq 0$ \cite{rque110}. This is confirmed in Fig. \ref{fig:smallXi}.
On the other hand, the rather rough ch$_0$ picture significantly improves the agreement with Monte Carlo 
data,
while the extended ch$_6$ description fares remarkably well (see Fig. \ref{fig:smallXi}). Extensive simulations 
have also been performed for larger $\Xi$ values, confirming the accuracy of the ch$_6$ route
for all values of the dielectric jump $\Delta$, while the simple ch$_0$ description is also shown to be quite accurate. 
In view of the underlying physical hypothesis (such as the two dimensional assumption for the
fluid of counter-ions), better justified for strongly coupled systems, the very good 
agreement at $\Xi=10$ rather comes as a surprise. A similar remark holds for ch$_0$, a crude,
but nonetheless trustworthy approximation.
It is interesting to compare and contrast our theory  with the approach of Reference \cite{BAO04} which also
also relies on the idea of singling out a test particle. However, at variance with our treatment,
a) the remaining
ions are treated at the Poisson-Boltzmann level ; b) the approach is restricted to $\Delta=0$,
and thus to a regime where many-body effects are less pronounced; c) the numerical resolution of a highly non-linear 
partial differential equation
is required, with subsequent numerical integration of some auxiliary potential.
In contrast, our treatment is fully analytical, and reduces to three simple
equations presented above.

\begin{figure}[htb]
\begin{center}
\includegraphics[width=0.42\textwidth,clip]{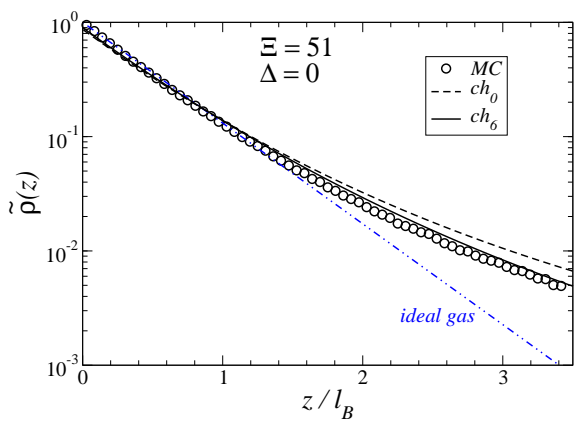}
\caption{Same as Fig. \ref{fig:smallXi}, without dielectric mismatch ($\Delta=0$),
and $\Xi=51$. The density profile is maximum for $z=0$, at contact with the plate:
there is no depletion zone ($z^*=0$).}
\label{fig:Delta0} 
\end{center}
\end{figure}

It is of particular interest to analyze the well documented $\Delta=0$ situation, where 
$\varepsilon=\varepsilon'$. There, the ideal gas view provides the dominant large coupling profile
\cite{MoNe02,NJMN05,SaTr11}. As seen in Fig. \ref{fig:Delta0}, both ch$_0$ and ch$_6$ 
perform significantly better, and account correctly for the deviations from the exponential behavior:
the overpopulated tail with respect to exponential behavior 
is a fingerprint of the repulsive effect of the fellow counter-ions
forming a layer at $z\simeq 0$, that becomes more pronounced as the test particle moves away 
from this plane. We have found a similar agreement at $\Delta=0$ for larger $\Xi$ values.

\begin{figure}[htb]
\begin{center}
\includegraphics[width=0.42\textwidth,clip]{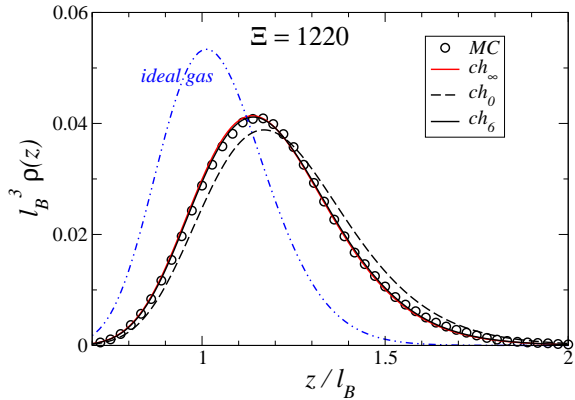}
\caption{Counter-ion profile at large coupling for $\Delta=1$, symbols are the results of MC simulations.
The ``Wigner strong coupling''
prediction (ch$_\infty$) is also shown: it is almost indistinguishable from the ch$_6$ treatment.}
\label{fig:largeXi} 
\end{center}
\end{figure}

Finally, we have tested our approach at very large couplings ($\Xi>10^3$), see Fig 
\ref{fig:largeXi}. While the ideal gas picture of Refs. \cite{MoNeEPL02,JKNP08}
is inoperative, the ch$_6$ theory agrees well with the simulation data, in spite
of the fact that the fluid of counter-ions is strongly modulated. We thus have considered extensions 
of ch$_6$, of the ch$_n$ type, including a growing number of neighbors in the approach
($n=6, 12, 18, 30\ldots$), that we locate at their ground state position, in order to reach 
gradually the $\Xi\to\infty$ hexagonal arrangement. Pushing this logic, we show in Fig. 
\ref{fig:largeXi} the ch$_\infty$ prediction, where all ions are in their ground state position,
except the test particle. It is still possible to compute analytically the resulting one body potential $u$
making use of the lattice summation techniques developed in Ref. \cite{SaTr12}.
There is barely any difference between the ch$_6$ and the ch$_\infty$ 
predictions. Incidentally, all ch$_n$ formulations, for $n$ between 6 and $\infty$,
remain extremely close for all
couplings we have investigated, which emphasizes the robustness of the approach \cite{rque64}. 
Furthermore, the depletion zone extension,
$z^*$, hardly depends on the level $n$ in a ch$_n$ treatment, from $n=0$ up to $n\to\infty$!

\section{Conclusion}

In conclusion, we have presented a theory that accounts very accurately for the ionic density profiles of salt-free 
systems at moderate and strong couplings. 
Extensive comparisons with Monte Carlo simulations have been carried out. 
Our approach is accurate
for $\Xi>10$, and thus covers a wealth of experimentally relevant
situations; for instance, DNA with trivalent counter-ions ($q=3$) has $\Xi$ around $100$. 
The couplings that both evade mean-field and our analysis, namely $\Xi$ in the range $[1,10]$,
must be addressed by computer simulation. 
Our formulation relies on basic electrostatics considerations, at variance with other
more complex treatments such as the splitting field-theory \cite{Chen06,Sant06,HaLu10,LuLi15},
and invokes transparent physical hypothesis pertaining to ionic correlations.
The latter are accounted for at a one body level, which qualifies the approach as mean-field.
Furthermore, besides accuracy, our treatment has been shown to be very robust.
More complex geometries such as a slit, explored for  small
separations in Ref. \cite{JKNP08} 
provide possible applications for the theory presented in this paper.

Another important perspective includes addition of co-ion \cite{persp}, which brings
an extra coupling parameter and hard core effects. This leads to significant complications,
but can elaborate on the no-salt treatment presented here, in the spirit of previous
approaches \cite{KNFP11,PaTr11}. On general grounds, salt ions "dress" the interactions
between multivalent counterions \cite{KNFP11}, in a way that may be complex, but that may admit rather
simple limiting laws. For instance, with highly asymmetric electrolytes,
counter-ions may be in a strong coupling regime while coions are not. This leads to 
a picture where the counterions interact 
through a screened potential, which allows further progress \cite{KNFP11}. 
Alternatively, if coions themselves are stronly coupled, 
they will form Bjerrum pairs with the counter-ions, leading to a system with excess
counter-ions and a number of dipoles, see e.g. \cite{Yan2011}. In a first approximation,  neglecting from pairs \cite{PaTr11}, 
the formalism presented here is directly applicable.

The support received from the Grant VEGA No. 2/0015/15 and from CNPq, INCT-FCx, and  US-AFOSR under the grant 
FA9550-16-1-0280 is acknowledged.



\begin{thebibliography}{11}
\bibitem{rque20}
When $\Delta > 0$, $\mu' \neq \mu$; it can be shown that 
$\mu' \propto \mu \,\Xi^{1/4}$ at large $\Xi$ \cite{SaTr12CPP}.
\bibitem{rque90}
This leaves a $\Xi$-gap between 1 and 10 where no analytical theory
has been proposed, while $\Xi<1$ is covered by the Poisson-Boltzmann theory.
\bibitem{rque50}
Each ion is surrounded by a first shell of nearest neighbors
which defines the coordination number in the liquid.
Interestingly, this coordination number is very close
in the liquid and crystalline phases \cite{DEG12}.
\bibitem{rque97}
In Ref. \cite{BdSL11}, both steps were merged, resulting in a double-layer
where all counter-ions move coherently, which is unphysical. It is important to disentangle
the calculation of the optimal distance $z^*$ from that of the effective
one-body potential $u$.
\bibitem{rque60}
In line with the argument given in the caption of Fig. \ref{fig:model},
$\sigma a^2 \propto q$. The prefactor $3^{-1/4} \sqrt{2}$
is chosen so that $a$ coincides with the lattice constant in the 
crystalline phase, a triangular (hexagonal) arrangement.
The $a$ thus defined also provides an excellent measure of
the nearest neighbor typical distance in the liquid phase \cite{MoNe02}.
\bibitem{rque70}
We sum the interactions between the test particle and its 6 nearest neighbors at 
a distance $\sqrt{(z-z^*)^2+a^2}$, their images at a distance 
$\sqrt{(z+z^*)^2+a^2}$, the particle self-image at a distance $2z^*$,
and the two punctured plates at $\pm z^*$, having hole radius $R_6$.
\bibitem{rque110}
A more precize criterion for the validity of the ideal gas picture is
$z^* \ll a$. This can be met for small enough $\Delta$ at large coupling,
but is incorrect for $\Delta$ above 0.3.
\bibitem{rque64}
This is a distinctive feature compared to previous approaches such as 
Ref. \cite{BdSL11}, where increasing the number of discrete neighbors results in poor
predictions.
\end{thebibliography}
\end{document}